\documentclass[prd,aps,floats,preprintnumbers,preprint]{revtex4}

\usepackage{graphicx}
 
\newcommand{\be}{\begin{equation}}
\newcommand{\ee}{\end{equation}}
\newcommand{\bea}{\begin{eqnarray}}
\newcommand{\eea}{\end{eqnarray}}

\begin{document}


\setlength{\unitlength}{1mm}

\title{Conditions for low-redshift positive apparent acceleration in smooth inhomogeneous models}

\author{Antonio Enea Romano$^{1,2,3}$}
\affiliation{
${}^1$Instituto de Fisica, Universidad de Antioquia, A.A.1226, Medellin, Colombia\\
${}^2$Department of Physics, National Taiwan University, Taipei 10617, Taiwan, R.O.C.\\
${}^3$Leung Center for Cosmology and Particle Astrophysics, National Taiwan University, Taipei 10617, Taiwan, R.O.C.\\
${}^4$Yukawa Institute for Theoretical Physics, Kyoto University,
Kyoto 606-8502, Japan
}

\begin{abstract}
It is known that a smooth LTB model cannot have a positive apparent central acceleration.  
Using a local Taylor expansion method we study the low-redshift conditions to obtain an apparent negative deceleration parameter $q^{app}(z)$  derived  from the luminosity distance $D_L(z)$ for a central observer in a LTB space, confirming that central smoothness implies a positive central deceleration.
Since observational data is only available at redshift greater than zero we find the critical values of the parameters defining a centrally smooth LTB model which give a positive apparent acceleration at $z>0$, providing a graphical representation of the conditions in the $q_0^{app},q_1^{app}$ plane, which are respectively the zero and first order terms of the central Taylor expansion of $q^{app}(z)$. We finally derive a coordinate independent expression for the apparent deceleration parameter based on the expansion of the relevant functions in red-shift rather than in the radial coordinate.

We calculate $q^{app}(z)$ with two different methods to solve the null geodesic equations, one based on a local central expansion of the solution in terms of cosmic time and the other one using the exact analytical solution in terms of generalized conformal time. 
 
\end{abstract}

\maketitle
\section{Introduction}

High redshift luminosity distance measurements require a dominant dark energy component  if interpreted under the assumption of inhomogeneity and isotropy of FLRW cosmological models.  
Because of the unknown physical nature of dark energy and the problems in the disagreement between the quantum field theory prediction of the value of the vacuum energy compared to the value of the cosmological constant inferred from these observations, some  alternatives have been 
proposed, such as the possibility that we may be located \cite{Nambu:2005zn,Kai:2006ws}
 at the center of an inhomogeneous isotropic universe 
described by a Lemaitre-Tolman-Bondi (LTB) solution of Einstein's field 
equations.
A  general approach to map the luminosity distance as a function of
 redshift $D_L(z)$ to LTB models has been 
 developed in \cite{Chung:2006xh},
 showing that an inversion method can be applied successfully to 
reproduce the observed $D_L(z)$.   
Interesting analyses of inhomogeneous cosmological models and other related problems are given for example in \cite{Romano:2006yc,Alexander:2007xx,Alnes:2005rw,GarciaBellido:2008nz,GarciaBellido:2008gd,GarciaBellido:2008yq,February:2009pv,Uzan:2008qp,Quartin:2009xr,Quercellini:2009ni,Clarkson:2007bc,Zuntz:2011yb,Ishibashi:2005sj,Bolejko:2011ys,Valkenburg:2011ty,deLavallaz:2011tj,Clifton:2009kx,Zibin:2011ma,Bull:2011wi,Clifton:2008hv,Zumalacarregui:2012pq,Marra:2011ct,Yoo:2010qn,Yoo:2008su,Clarkson:2010uz,Kodama:2010gr,Clarkson:2012bg,Kroupa:2012qj,Bolejko:2012ue,Valkenburg:2012ds,Clifton:2012qh}.
The idea to use galaxy number counts \cite{Romano:2007zz} to distinguish 
between inhomogeneous models  and $\Lambda$CDM has been 
 investigated both analytically  and 
numerically \cite{Romano:2009qx,Romano:2009ej,Celerier:2009sv} , showing how not centrally smooth LTB models can mimick the effects of a cosmological constant both for 
the redshift spherical shell mass $mn(z)$ and $D_L(z)$.
More recently there has been clear evidence that LTB solutions cannot provide a fully consistent cosmological model
compatible with all available observations \cite{Moss:2011ze,Zhang:2010fa}.
They nevertheless provide a good toy model to understand the effects of inhomogeneities which motivate the present study.

In this paper we compute the low redshift expansion of the apparent deceleration parameter $q^{app}(z)$ for an observer located at the center of a centrally smooth LTB model.
After confirming that the central value of the apparent acceleration $q^{app}(0)$ is positive, we derive the general conditions for a negative $q^{app}(z_n)$ at some arbitrarily low red-shift $z_n$.
This is motivated by the fact that observational data is only available at redshift greater than zero, but still sufficiently small to make the Taylor expansion accurate.

We give a graphical representation of the conditions for a negative apparent deceleration in the $(q_0^{app},q_1^{app})$ plane, which are respectively the zero and first order terms of the central Taylor expansion of $q^{app}(z)$.
We calculate $q^{app}(z)$ with two different methods to solve the null geodesic equations: one based on a local central expansion of the solution of the Einstein's equations in terms of cosmic time, reported in the Appendix, and another one using the exact analytical solution, confirming the two methods are in agreement.

The expansion of the solution in terms of cosmic time is quite useful also for other applications requiring a foliation of space-time in space-like hyper-surfaces, such as spatial averaging, which is much more difficult to study using the analytical solution in terms of generalized conformal time coordinate.
  
\section{The Lemaitre-Tolman-Bondi (LTB) metric}
Lemaitre-Tolman-Bondi  solution can be
 written as \cite{Lemaitre:1933qe,Tolman:1934za,Bondi:1947av}
\begin{eqnarray}
\label{eq1} %
ds^2 = -dt^2  + \frac{\left(R,_{r}\right)^2 dr^2}{1 + 2\,E}+R^2
d\Omega^2 \, ,
\end{eqnarray}
where $R$ is a function of the time coordinate $t$ and the radial
coordinate $r$, $R=R(t,r)$, $E$ is an arbitrary function of $r$, $E=E(r)$
and $R,_{r}=\partial R/\partial r$.

The first integral of Einstein's equations give
\begin{eqnarray}
\label{eq2} \left({\frac{\dot{R}}{R}}\right)^2&=&\frac{2
E(r)}{R^2}+\frac{2M(r)}{R^3} \, , \\
\label{eq3} \rho(t,r)&=&\frac{2 M,_{r}}{R^2 R,_{r}} \, ,
\end{eqnarray}
with $M=M(r)$ being an arbitrary function of $r$ . We are denoting with a dot 
the partial derivative with respect to $t$, $\dot{R}=\partial R(t,r)/\partial t$.
There analytical solutions of Eq.\ (\ref{eq2})  expressed  
in terms of a time variable $\tau=\int^t dt'/R(t',r) \,$ is
\begin{eqnarray}
\label{eq4} Y(\tau ,r) &=& \frac{M(r)}{- 2 E(r)}
     \left[ 1 - \cos \left(\sqrt{-2 E(r)} \tau \right) \right] \, ,\\
\label{eq5} t(\tau ,r) &=& \frac{M(r)}{- 2 E(r)}
     \left[ \tau -\frac{1}{\sqrt{-2 E(r)} } \sin \left(\sqrt{-2 E(r)}
     \tau \right) \right] + t_{b}(r) \, ,
\end{eqnarray}
where  $Y$ has been introduced to make clear the distinction
 between the two functions $R(t,r)$ and $Y(\tau,r)$
 which are trivially related by 
\begin{equation}
R(t(\tau,r))=Y(\tau,r) \, ,
\label{Rtilde}
\end{equation}
and $t_{b}(r)$ is another arbitrary function of $r$, which plays the role of a functional constant of integration and is commonly called the bang function,
corresponding to the the possibility of inhomogenous big-bang/crunches times. 


After introducing the functions
\begin{equation}
 A(t,r)=\frac{R(t,r)}{r},\quad k(r)=-\frac{2E(r)}{r^2},\quad
  \rho_0(r)=\frac{6M(r)}{r^3} \, ,
\end{equation}
so that  Eq.\ (\ref{eq1}) and the Einstein equations
(\ref{eq2}) and (\ref{eq3}) are written in a form 
similar to those for FLRW models,
\begin{equation}
\label{eq6} ds^2 =
-dt^2+A^2\left[\left(1+\frac{A,_{r}r}{A}\right)^2
    \frac{dr^2}{1-k(r)r^2}+r^2d\Omega_2^2\right] \, ,
\end{equation}
\begin{eqnarray}
\label{eq7} %
\left(\frac{\dot{A}}{A}\right)^2 &=&
-\frac{k(r)}{A^2}+\frac{\rho_0(r)}{3A^3} \, ,\\
\label{eq:LTB rho 2} %
\rho(t,r) &=& \frac{(\rho_0 r^3)_{, r}}{3 A^2 r^2 (Ar)_{, r}} \, .
\end{eqnarray}
The solution of Eqs.\ (\ref{eq4}) and (\ref{eq5}) is now written as
\begin{eqnarray}
\label{LTB soln2 R} a(\eta,r) &=& \frac{\rho_0(r)}{6k(r)}
     \left[ 1 - \cos \left( \sqrt{k(r)} \, \eta \right) \right] \, ,\\
\label{LTB soln2 t} t(\eta,r) &=& \frac{\rho_0(r)}{6k(r)}
     \left[ \eta -\frac{1}{\sqrt{k(r)}} \sin
     \left(\sqrt{k(r)} \, \eta \right) \right] + t_{b}(r) \, ,
\end{eqnarray}
where $\eta \equiv \tau\, r = \int^t dt'/A(t',r) \,$ and $A(t(\eta,r),r)=a(\eta,r)$.
From the above definitions we can consider $\eta$ a generalized conformal time coordinate.

Due to the freedom in the choice of the radial coordinate we can set 
the function $\rho_0(r)$ to be a constant,
 $\rho_0(r)=\rho_0=\mbox{constant}$, corresponding to the choice of coordinates in which $M(r)\propto r^3$, the so called \cite{Tanimoto:2007dq} the FLRW gauge.
We will adopt these coordinates in the rest of this paper. 
\section{Low redshift expansion}
Since the geodesic equations are simpler if written in in terms of the variable $t$, it can be useful to find a perturbative solution of for $R(t,r)$, expanding the Einstein's equations around the point $(t_0,0)$ corresponding to the central observer.

We will expand the relevant functions in the following way:

\bea
t(z)&=&t_0+t_1 z+t_2 z^2+t_3z^3, \\
r(z)&=&r_0+r_1 z+r_2 z^2+r_3 z^3 , \label{rz}\\
E(r)&=&e_2 r^2+e_4 r^4, \\
k(r)&=&k_0+k_2 r^2 \,, \label{kr}
\eea
where we have only even powers for $E(r)$ and $k(r)$ to ensure that the solution is analytical everywhere. 
In order to obtain a perturbative expansion for the solution it is convenient to introduce a new variable $x$ and a function $f(x)$ according to:
\bea
R=\frac{M}{|2E|}f(x) ,\\
x=\frac{|2E|^{3/2}t}{M}=\frac{|k|^{3/2} t}{\rho_0},
\eea
in terms of which the Einstein's equations corresponds to:
\bea
p=\frac{|E|}{E} \\
f'(x)^2-\frac{2}{f}-p=0.
\label{eqf}
\eea
This choice of variables is different from \cite{Tanimoto:2007dq}, and in fact it leads to the simpler differential equation (\ref{eqf}). Another difference is that the relations derived below in eq.(\ref{f1}-\ref{f3}) for the coefficients of the expansion are
valid for any sign of $k(r)$ , while in \cite{Tanimoto:2007dq} a two branches solution is used, requiring the necessity of the additional proof that the expansion is the same for each branch.

The main advantage of re-writing the solution in this form is that we now have a constant coefficients differential equation in the single variable $x$, which is easier to Taylor expand.

We can then solve this differential equation perturbatively around $x_0=x(t_0,0)$
\bea
f(x) &=& f_0+f_1(x-x_0)+\frac{f_2}{2!}(x-x_0)^2+\frac{f_3}{3!}(x-x_0)^3+\frac{f_4}{4!}(x-x_0)^4,\\ 
f_1=\sqrt{\frac{2}{f_0}+p} && f_2=-\frac{1}{f_0^2} ,\\
f_3=\frac{2 \sqrt{\frac{f_0 p+2}{f_0}}}{f_0^3} && f_4=-\frac{2 (3 f_0 p+7)}{f_0^5}.
\eea
The above expansion of the solution has been obtained without any knowledge of the exact anaytical solution, but combining this latter with the definitions of $f(x)$ we obtain:

\bea
f(y)=-p(1-\cos{y}), \label{f1} \\  
x(y)=(-p){}^{3/2}(y-\sin{y}), && y=\sqrt{k} \eta, \label{f2} \\
f_0=f(x_0)=-p_0(1-\cos{\sqrt{k_0}\eta_0}), && p_0=\frac{|e_2|}{e_2}=-\frac{|k_0|}{k_0}, \label{f3}
\eea
which can be used to derive with an alternative method the coefficients of the expansion of $f(x)$, and in particular to fix $f_0$. 
We can observe that $f_0$, which encodes the initial conditions of the solution since all the other $f_i$ coefficients depend on it, depends only on the product $\sqrt{k_0}\eta_0$, implying a re-scaling symmetry on $\eta_0$ and $k_0$ under which observational quantities are invariant.
Applying recursively the derivative chain rule we can for example obtain the linear and second order terms using :
\bea
\frac{\partial f}{\partial x}=\frac{\partial f}{\partial y}\frac{\partial y}{\partial x}=\frac{\partial f}{\partial y}\left(\frac{\partial x}{\partial y}\right)^{-1} ,\\
\frac{\partial^2 f}{\partial x^2}=\frac{\partial }{\partial y}\left(\frac{\partial f}{\partial y}\left(\frac{\partial x}{\partial y}\right)^{-1}\right)\left(\frac{\partial x}{\partial y}\right)^{-1} .\\
\eea
It can be verified that the coefficients $f_i$ obtained solving perturbatively the differential equation (\ref{eqf}) or  using the above mentioned method are the same. These relations are also important to establish the a connection with the parameters appearing in the exact solution.

In order to check the equivalence of the calculations done using the two different set of coordinates $(\eta,r)$ and $(t,r)$, it is also useful to define

\be
t_0=t(\eta_0,0) = \frac{\rho_0}{6k_0}\left[ \eta_0 -\frac{1}{\sqrt{k_0}}\sin\left(\sqrt{k_0} \, \eta_0 \right) \right] + t_{b}(0) \, . \label{t0}
\ee
\section{Geodesics equations and luminosity distance}
The geodesics equations are \cite{Celerier:1999hp}:
\begin{eqnarray}
{dr\over dz}={\sqrt{1+2E(r(z))}\over {(1+z)\dot {R'}[r(z),t(z)]}} \,,
\label{eq:34} \\
\nonumber
{dt\over dz}=-\,{R'[r(z),t(r)]\over {(1+z)\dot {R'}[r(z),t(z)]}} \,, 
\label{geotr} \\
\end{eqnarray}
where the $'$ denotes the derivative respect to $r$ and the dot $\dot{}$ the derivative respect to $t$. 
These equations are derived from the definition of redshift and by following the evolution of a short time interval along the null geodesic $T(r)$.
The r.h.s. can be evaluated from the perturbative solution for $R(t,r)$ in terms of $f(x)$.
Expanding in powers of $z$ we can then reduce the solution of this system of partial differential equations to solving a system of linear algebraic equations, where the unknowns are the expansion coefficients:
\bea
G(z)={dr\over dz}=G_0+G_1 z+G_2 z^2,\\
L(z)={dt\over dz}=L_0+L_1 z+L_2 z^2,\\
r(z)=r_0+r_1 z+r_2 z^2+r_3 z^3,\\
t(z)=t_0+t_1 z+t_2 z^2,\\
r_1=G_0 , 2 r_2= G_1 , 3 r_3=G_2 , t_1=L_0 , 2 t_2=L_1.
\eea
We will assume also assume a homogeneous big-bang, $t_b(r)=0$
In the above equations we are expanding all the quantities to the order necessary to expand the luminosity distance to the third order.
For the geodesic equations we get:
\bea
   r(z)&=&\sqrt{\frac{{f_0} }{{2} {e_2 p_0} (f_0 p_0 +2)}}z-\frac{\sqrt{f_0}  (f_0 p_0+3)}{2 \sqrt{2 e_2 p_0} (f_0p_0+2)^{3/2}} z^2+\\
   &&+ \frac{1}{24e_2^3 \rho_0 (f_0 p_0+2)^3}\bigg[ e_2^2 \left(2\rho_0     \sqrt{2 f_0 e_2 p_0 (f_0 p_0+2)}     \left(3 f_0^2 p_0+12 f_0+15p_0\right)+216 e_4 t_0 (f_0 p_0+2)\right) \nonumber \\
   && -3 \sqrt{2} e_4 f_0^{3/2}  \sqrt{e_2 p_0} (f_0 p_0+2)^{3/2}\bigg]z^3 ,\nonumber \\
t(z)&=&t_0-\frac{f_0^{3/2} \text{$\rho$0} \sqrt{e_2 p_0}}{12 \sqrt{2} e_2^2 \sqrt{f_0 p_0+2}}z +\frac{f_0^{3/2}  \rho_0 \sqrt{e_2 p_0} (2 f_0 p_0+5)}{24 \sqrt{2} e_2^2 (f_0 p_0+2)^{3/2}} z^2. 
\eea

Using eqs.(\ref{f1}-\ref{f3},\ref{t0}) it is more elegant and convenient to re-express our results in terms of $\eta_0$ ,$k_i$, and $\rho_0$ which gives the following formula for the luminosity distance $D_L(z)$  :

\bea
D_L(z)&=&(1+z)^2R(t(z),r(z))=(1+z)^2r(z)^2a(\eta(z),r(z)) ,\\
D_L(z)&=&D^1_L z+D^2_L z^2+D^3_L z^3+ ... \nonumber \\
D^1_L&=&\frac{B^3 \rho_0}{k_0^{3/2} \left(3 B^2+3\right)} ,\nonumber \\
D^2_L&=&-\frac{B^3 \left(B^2-1\right) \rho_0}{12 k_0^{3/2} \left(B^2+1\right)} ,\nonumber \\
D^3_L&=&\frac{B^2 \rho_0 \left(k_0^2  \left(B^4-1\right)+8 k_2 B^2 X+2 k_2 \left(3 B^4 X-9 B^3+8 B^2 X-9 B+9 X\right)\right)}{24 k_0^{7/2}
   \left(B^2+1\right)} ,\nonumber \\
X &= \sqrt{k_0}\eta_0/2 ; \;\; & B=\tan{(X)} .
\eea

In the rest of the paper we will not give the formulae in terms of $\eta_0$ and trigonometric functions, since they are rather complicated. So far we have in fact worked in terms of functions and parameters which depend on the coordinate choice, but since they are not directly observable quantities, it is more convenient to introduce :

\bea
a_0=\frac{B^2 \rho_0}{3 k_0 (B^2+1)}, \\
H_0=\frac{3 k_0^{3/2} \left(B^2+1\right)}{B^3 \rho_0}, \\
q_0=\frac{1}{2} \left(B^2+1\right), \label{q0v}
\eea
where we have used the following definitions 
\bea
a_0&=&a(\eta_0,0)\,, \\
H_0 & = &\frac{\dot{a}(t_0,0)}{a(t_0,0)} \,,\\
q_0 &= -&\frac{\ddot{a}(t_0,0)\dot{a}(t_0,0)}{\dot{a}(t_0,0)^2} \label{q0}\,.
\eea

From the equations\ for $q_0$ we can see that it must be always positive. As we will see later, under the assumption of central smoothness we have adopted we have that $q_0=q_0^{app}$, in agreement with previous studies \cite{Vanderveld:2006rb}, which proved that the apparent deceleration parameter cannot be negative for a smooth LTB solution.

The formula for the luminosity distance takes now the form:
\bea
D^1_L&=&\frac{1}{H_0} ,\\
D^2_L&=&\frac{1-q_0}{2 H_0}, \nonumber \\
D^3_L&=&\frac{(q_0-1) q_0}{2 H_0}+\frac{3 K_2 q_0 \left[2 (q_0+1) \frac{\arctan\left(\sqrt{2 q_0-1}\right)}{\sqrt{2 q_0-1}} -3\right]}{2 H_0 (2 q_0-1)^2} ,\nonumber
\eea
where we have introduced the dimensionless parameter  $K_2=k_2/(a_0 H_0)^4$. The inhomogeneities effects show only from the third order coefficient $D^3_L$ because we have not included odd powers in the expansion of $k(r)$ in order to satisfy smoothness conditions, and in the $\{K_2=0,q_0=1/2\}$ limit we recover the case of a flat matter dominated FLRW Universe, i.e. $\Omega_M=1$.

From the definitions of $H_0$ , $q_0$ and $a_0$ we get the following relations with the parameters appearing in the exact solution:

\bea
X&=&\arctan{\sqrt{2q_0-1}} ,\\
k_0&=& a_0^2 H_0^2 (2 q_0-1) ,\label{k0} \\
\rho_0&=&6 a_0^3 H_0^2 q_0 .\label{rho0}
\eea

These formulas are valid for both negative and positive $k_0$ by analytical continuation, and provide a very compact and insightful picture of the physics of the problem.
Eq.(\ref{k0}) gives in fact the direct relation between $q_0$ and the value of $k_0$, and implies that $q_0>1/2$ only for positive $k_0$.

\section {Apparent cosmological observables}
The following relations apply to a flat FLRW model 
\bea
H^{app}(z)=\left[\frac{d}{dz}\left(\frac{D_L(z)}{1+z}\right)\right]^{-1}\,,
\\
Q^{app}(z)=\frac{d}{dz}\left(\frac{D_L(z)}{1+z}\right)=(H^{app}(z))^{-1}\,,
\\
q^{app}(z)=-1-\frac{d {\ln}(Q^{app}(z))}{d{\ln}(1+z)}=q^{app}(D_L(z))\,, \label{qapp}
\eea
Apparent observables are then defined as those obtained by applying the same relations above to the luminosity distance
$D^{LTB}_L(z)$ obtained for a central observer in a LTB model.
Another investigation of the effects of inhomogeneities on apparent cosmological observables in presence of a cosmological constant can be found in \cite{Romano:2010nc}, where is shown that the inhomogenity could introduce the illusion of a red-shift dependent equation of state, instead of a cosmological constant.
It can be easy verified that
\bea
q^{app}(0)=q^{app}_0&=&q_0 ,\;\label{q0appv}\\
H^{app}(0)=H^{app}_0&=&H_0,
\eea
where it is important to observe that $q_0$ and $q^{app}_0$ are  defined independently in eq. (\ref{q0}) and eq.(\ref{qapp}) respectively.
The equality between $q_0$ and $q^{app}_0$ only holds under the assumption that $k'(0)=t_b'(0)=0$ as shown in \cite{Romano:2009mr}.
This can be considered a natural consequence of the fact that for a smooth LTB model the apparent deceleration parameter $q_0^{app}$ at the center should coincide, by continuity, with  $q_0$.
As expected, in the centrally smooth case we are considering $q_0^{app}$ is always positive because of eq.(\ref{q0v}). This is an alternative \cite{Vanderveld:2006rb} and easier prove of the fact that a smooth LTB space has a positive central  apparent deceleration parameter:
\bea
q^{app}_0 &\geq&0 \,.
\eea
After substituting the expansion of the luminosity distance in the definition of apparent cosmological deceleration given in eq.(\ref{qapp})  we get
\bea
q^{app}(z)&=&q^{app}_0+q^{app}_1 z + . . .\\
q^{app}_0&=&\frac{1}{2} \left(B^2+1\right)=q_0 ,\\
q^{app}_1&=&-\frac{k_0^2 \left(B^2-1\right) B^3+9 k_2 \left(B^2+1\right) \left(B^2 X-3 B+3 X\right)}{2 k_0^2 B} .
\eea

As it can be seen, since we consider a centrally smooth model, the effects of the inhomogeneity show only in the linear order coefficient $q^{app}_1$, but this will be enough to derive the general leading order conditions for a negative deceleration at a red-shift different  from zero.

\section{Conditions for a negative apparent cosmological deceleration}
From the expression derived in the previous section we can deduce the leading order condition to have a negative $q^{app}(z_n)<0$ at a given redshift $z_n$ 
\bea
q^{app}_1&<&-\frac{1}{z_n}q^{app}_0 \label{qappcon} \,.
\eea

\begin{center}
\begin{figure}[h]
\includegraphics[height=60mm,width=80mm]{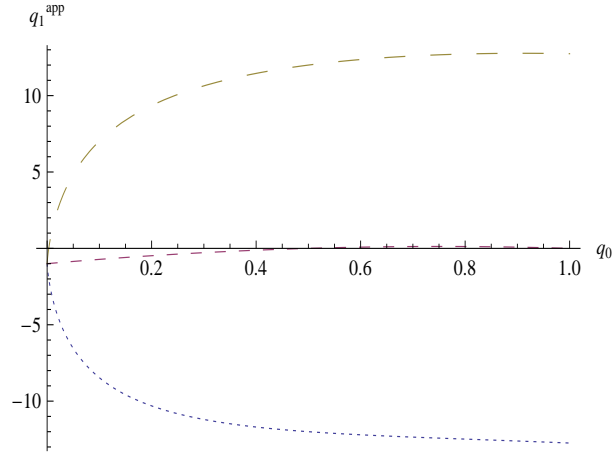}
\caption{ $q_1^{app}(q_0)$ is plotted for different values $K_2=k_2/(a_0^4 H_0^4)$ and for $0<q_0<1$. The small dashed line corresponds to $K_2=10$, the medium dashed line corresponds to $K_2=0$ and the long dashed line corresponds to $K_2=-10$. }
\label{q1k10}
\end{figure}
\end{center}

\begin{center}
\begin{figure}[h]
\includegraphics[height=120mm,width=120mm]{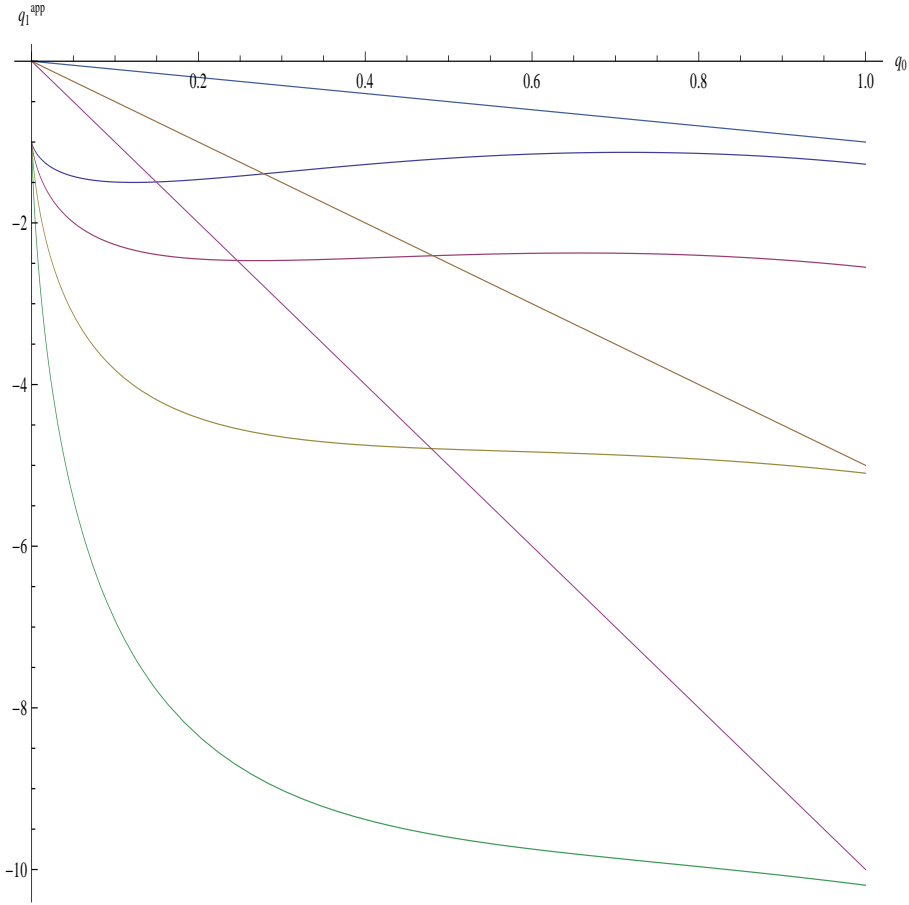}
\caption{ $q_1^{app}(q_0)$ is plotted for different positive values of $K_2=k_2/(a_0^4 H_0^4)=1,2,4,8$ and for $0<q_0<1$. The straight lines correspond to $q_1^{app}=-\frac{1}{z} q_0$ for different values of z, i.e. with different slopes. Negative values of $q^{app}(z_n)$ correspond, from eq.(\ref{qappcon}), to the area of the $(q_0,q^{app}_1)$ plane below the corresponding straight line with equation $q^{app}_1=-\frac{1}{z_n} q_0$.
This implies that, keeping $K_2$ fixed, for smaller redshifts, i.e lower slope lines, the maximum value of $q_0$ which gives a negative $q^{app}(z)$ decreases. We can also see that at the some redshift, i.e. for the same straight line, higher values of $K_2$ give a larger maximum value of $q_0$ necessary to satisfy the condition for a negative $q^{app}(z)$. Increasingly Higher values of $K_2$ correspond to curves with increasingly lower values of $q_1^{app}(q_0)$.}
\label{q1kAll}
\end{figure}
\end{center}
Since $q^{app}_0$ is always positive, we will focus here on the linear order correction, and after introducing the dimensionless parameter $K_2=k_2/(a_0^4 H_0^4)$ and using eq.(\ref{q0appv}), we can express the first order coefficient in the following form:
\be
q^{app}_1=(1-q_0) (2
   q_0-1)-\frac{9 K_2 q_0 \left(2 (q_0+1) \arctan\left(\sqrt{2 q_0-1}\right)-3 \sqrt{2 q_0-1}\right)}{(2 q_0-1)^{5/2}}.
\ee
From this last formula we can see that the only physically relevant parameters are the dimensionless quantities $q_0$ and  $K_2$.
From the figs. 1-2 representing $q_1^{app}(q_0)$ we can see that larger $K_2$ allow $q^{app}(z)$ to be negative for smaller maximum values of $q_0$. Because of eq.(\ref{k0}) $q_0$ and $K_0$ are in fact proportional, and so a smaller value of $K_0$  is compensated by a faster growth of $k(r)$, which is the same reason why negative $K_2$, corresponding to a decreasing $k(r)$, requires larger values of $q_0$.
The straight lines drawn in the fig. \ref{q1kAll} have slope $-\frac{1}{z}$, and from eq.(\ref{qappcon}) a negative $q^{app}(z)$ correspond to the area of the $(q_0,q^{app}_1)$ plane below those critical lines.
This implies that, keeping $K_2$ fixed, for smaller redshift, i.e lower slope lines, the maximum value of $q_0$ which gives a negative $q^{app}(z)$ decreases. From the same figure we can see that at the same redshift, i.e. for the same straight line, higher values of $K_2$ give a larger maximum value of $q_0$ necessary to satisfy the condition for a negative $q^{app}(z)$.

\section{Coordinate independent conditions}

So far we have derived the conditions for a negative apparent deceleration parameter under the choice of the radial coordinate
in which $M(r)=\frac{\rho_0 r^3}{6}$, but it would be more useful to derive a set of coordinate independent conditions.
This can be achieved by considering the expansion of the relevant functions in terms of the red-shift:
\bea
k(z)&=&k_0+k^z_1 z+ k^z_2 z^2+ ..
\eea
After substituting the expansion for $r(z)$ in $k(r)$ we get
\bea
k(z)&=&k(r(z))=k_0 + (k_1 r_1) z + \frac{1}{2} (2 k_2 r_1^2 + k_1 r_2) z^2 \,,
\eea
from which we obtain the relations
\bea
k_1&=&\frac{k^z_1}{r_1}  \,, \label{k1z}\\
k_2&=&-\frac{-2 k^z_2 r_1+k^z_1 r_2}{2 r_1^3}\,, \label{k2z}
\eea
which imposing the smoothness condition reduce to
\bea
k_1&=&0 \,, \\
k_2&=&\frac{k^z_2}{r_1^2} \,.
\eea
In the above expression the coefficients $k_i,r_i$ are the same as the ones defined in equations (\ref{rz},\ref{kr}).

We can now substitute the above relations in the formulae obtained in the previous section, finally getting an expression for $q^{app}(z)$ which only depends on the coordinate independent coefficient of the redshift expansion $K^z_i$:
\bea
q^{app}(z)&=&q_0^{app}+q_1^{app}z+..=\frac{1+K_0}{2}+\frac{1}{{2 K_0^{5/2}}}\bigg[ (\sqrt{K_0} (K_0^3-K_0^4+27 K^z_2+27 K_0 K^z_2)+ \nonumber \\
&&-9 (3+4 K_0+K_0^2) K^z_2 \arctan{\sqrt{K_0}}\bigg] z+ .. \,, \label{qappkz}
\eea
where we have expressed the final result in terms of the dimensionless parameters
\bea
K_0&=&k_0(a_0 H_0)^{-2}=(2q_0-1) \,, \\
K^z_2&=&k^z_2(a_0 H_0)^{-2} \,. 
\eea
The condition for a positive acceleration remains the same as in eq.(\ref{qappcon}).

It should be noted that while the expansion of $r(z)$ is different for different choices of the radial coordinate, that for for $k(z)$ is coordinate independent, implying that the above expression for $q^{app}(z)$ can be applied to any LTB model.
It can be checked explicitly in fact that for example adopting the light cone gauge the expansion for $r(z)$ would be different, but the formula (\ref{qappkz}) would be the same since it is a physical observable. 

The redshift expansion of $k(z)$ would also be the same since
\bea
k^{fr}[r_{fr}(z)]&=&k^{fr}[r_{fr}(r_{lc}(z)]=k^{lc}[r_{lc}(z)]=k(z) \,, \\
k^{lc}[r_{lc}]&=&k^{fr}[r_{fr}(r_{lc})] \,,
\eea 
where for clarity we are denoting with $r_{lc}$ the light cone gauge radial coordinate and $r_{fr}$ the FLRW gauge radial coordinate.
A direct check would require to derive explicitly a local expansion of the coordinate transformation between $r_{fr}(r_{lc})$ which can be obtained by imposing the condition:
\bea 
M^{lc}[r_{lc}(r_{fr})]&=&M^{fr}[r_{fr}]=\frac{\rho_0 r_{fr}^3}{6} \,, \\
r_{lc}(r_{fr})&=&r_{1}^{lc} r_{fr}+r_2^{lc} (r_{fr})^2+r_3^{lc} (r_{fr})^3 + ..
\eea
This is a rather cumbersome procedure, so we will not report it here, but simply observe that our final result in eq.(\ref{qappkz}) is coordinate independent because it involves the coefficients of the expansion of $k(z)$ and not of $k(r)$.
A graphical representation in the plane ${K_0,q_1^{app}}$ of the condition for a positive acceleration is give in Fig. 3. 

\begin{center}
\begin{figure}[h]
\includegraphics[height=60mm,width=80mm]{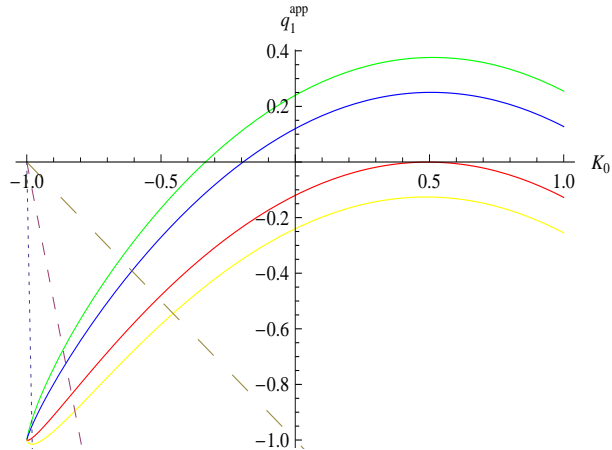}
\caption{ $q_1^{app}$ is plotted for different values of $K^z_2=\{-0.2,-0.1,0.1,0.2\}$ and for  $-1<K_0<1$. The small dashed line corresponds to $z=0.01$, the medium dashed line corresponds to $z=0.1$ and the long dashed line corresponds to $z=5$. Negative values of $q^{app}(z_n)$ correspond, from eq.(\ref{qappcon}), to the area of the $(K_0,q^{app}_1)$ plane below the corresponding straight lines with equation $q^{app}_1=-\frac{1}{z_n} q_0=-\frac{1}{z_n} \frac{1+K_0}{2}$.
Higher values of $K^z_2$ correspond to lower values of $q_1^{app}$.}
\label{q1k01}
\end{figure}
\end{center}

\section{Conclusions}

We have derived the low-redshift expansion for the apparent cosmological deceleration parameter $q^{app}(z)$ for a central observer at the center of a centrally smooth spherically symmetric matter inhomogeneity.
The same results have been derived with two different methods, one based on the existence of an exact solution in terms of coordinates $(\eta,r)$, and the other using a local perturbative expansion of the solution around the center using cosmic time.
We have then have applied it to study the conditions for a negative value of the apparent deceleration parameter $q^{app}(z)$ at non zero redshift, showing how the maximum allowed value of $q_0$  decreases as the redshift decreases, and it decreases as $K_2$ decreases. 
We have finally derived a coordinate independent expression for the apparent deceleration parameter, based on the red-shift expansion of the relevant functions.

The method we used to expand the solution of the Einstein's equation has the advantage of using cosmic time and not the parametric analytical solution, allowing to use it for other applications which require space-like foliation of space-time such as volume averaging.

In the future it will be interesting to apply our analytical methods to study other relevant cosmological observables for a central observers in $LTB$ spaces, in order to better understand the local effects effects of large scale inhomogeneities.

\begin{acknowledgments}
This work was also supported in part by JSPS Grant-in-Aid for Scientific 
Research (A) No.~21244033, and by JSPS 
Grant-in-Aid for Creative Scientific Research No.~19GS0219,
and by Monbukagaku-sho Grant-in-Aid for the global COE program,
"The Next Generation of Physics, Spun from Universality and Emergence".
A.E.R. thanks Misao Sasaki, Alexei Starobinsky and Diego Restrepo.
A.E.R. is also supported by the CODI of UDEA, and the dedicacion exclusiva program of the Vicerectoria de Docencia of UDEA.
\end{acknowledgments}

\appendix
\section {Alternative method}

Here we adopt the same method developed in \cite{Romano:2009xw} to find the null geodesic equation in the coordinates $(\eta,t)$, to  find a local expansion of the solution around $z=0$ corresponding to  $(t_0,0)\equiv(\eta_0,t)$, where $t_0=t(\eta_0,r)$.
The luminosity distance for a central observer in a LTB space can be written as 
\be
D_L(z)=(1+z)^2 R\left(t(z),r(z)\right)
=(1+z)^2 r(z)a\left(\eta(z),r(z)\right) \,,
\ee
where $\Bigl(t(z),r(z)\Bigr)$ or $\Bigl((\eta(z),r(z)\Bigr)$
is the solution of the radial geodesic equation
as a function of the redshift in different coordinates.

The  radial null geodesic equation is
\bea
\label{geo1}
\frac{dT(r)}{dr}=f(T(r),r) \,;
\quad
f(t,r)=\frac{-R_{,r}(t,r)}{\sqrt{1+2E(r)}} \,.
\eea
where $T(r)$ is the time coordinate along the null radial geodesic as a function of the the coordinate $r$.
From the analytical solution, we can write 
\bea
T(r)&=&t(U(r),r) \,, \\ 
\frac{dT(r)}{dr}&=&\frac{\partial t}{\partial \eta} \frac{dU(r)}{dr}+\frac{\partial t}{\partial r} \,,
\eea
where $U(r)$ is the $\eta$ coordinate along the null  geodesic as a function of the the radial coordinate $r$.
It is then possible to write the geodesic equations for the coordinates \cite{Romano:2009xw} $(\eta,r)$,
\bea
\label{geo3}
\frac{d \eta}{dz}
&=&\frac{\partial_r t(\eta,r)-F(\eta,r)}{(1+z)\partial_{\eta}F(\eta,r)}=p(\eta,r) \,,\\
\label{geo4}
\frac{dr}{dz}
&=&-\frac{a(\eta,r)}{(1+z)\partial_{\eta}F(\eta,r)}=q(\eta,r) \,, \\
F(\eta,r)&=&-\frac{1}{\sqrt{1-k(r)r^2}}\left[\partial_r (a(\eta,r) r)
+\partial_{\eta} (a(\eta,r) r) \partial_r \eta\right]  \, , 
\eea
where $\eta=U(r(z))$ and $F(\eta,r)=f(t(\eta,r),r)$.
It is important to observe that the functions $p,q,F$ have an explicit analytical form which can be obtained from $a(\eta,r)$ and $t(\eta,r)$.

In order to take advantage of the fully analytical expression of the equation of the radial null geodesics we can follow the same procedure we adopted for the coordinate $(t,r)$, and expand the geodesic equations solution as:
\bea
\eta(z)&=&\eta_0+\eta_1 z+\eta_2 z^2+ ... \\
\eea
The result for $\eta(z)$ is
\bea
\eta(z)&=&\frac{2 \arctan\left(\sqrt{2 q_0-1}\right)}{H_0 \sqrt{2
   q_0-1}}-\frac{z}{H_0}+ \\ \nonumber
&&+\frac{ \left(\sqrt{2 q_0-1} \left(H_0^4 \left(4 q_0^3-3 q_0+1\right)-3 k_2\right)+2 k_2 (q_0+1) \arctan\left(\sqrt{2
   q_0-1}\right)\right)}{2 H_0^5 (2 q_0-1)^{5/2}}z^2 \,,
\eea   
while  $r(z)$ is the same as the one obtained in the previous sections since the coordinate $r$ does not change.

It can be checked that the two methods to calculate the geodesics are equivalent by substituting in the exact solution $t(\eta,r)$
\be
t(z)=t(\eta(z),r(z)) \,,
\ee
the expressions $\eta(z)$ and $r(z)$ and comparing with the results for $t(z)$ from the previous sections.
Such a substitution gives exactly the some expression for $t(z)$, proving that the two methods to obtain the red-shift expansion of the solution of the geodesics equations are equivalent.

\end{document}